\documentclass[stef,twoside]{stefano}

\usepackage{amsmath}
\usepackage{amssymb}
\usepackage{graphics}
\usepackage{rotating}
\usepackage{cite}
\usepackage{color}

\def\makeheadbox{{%
\hbox to0pt{\vbox{\baselineskip=10dd\hrule\hbox
to\hsize{\vrule\kern3pt\vbox{\kern3pt \hbox{{\sc Eur. Phys. J. C} {\bf 62},
793-797 (2009) }
\hbox{
{\sc {\color{blue}{dma}}[{\color{black}{imecc}}]{\color{red}{UniCamp}}
}
\hspace*{10.3cm}
{\color{blue}{$\boldsymbol{\Sigma \delta \Lambda}$ }}}
\kern3pt}\hfil\kern3pt\vrule}\hrule}%
\hss}}}

\def\0{\mbox{\tiny $0$}}
\def\1{\mbox{\tiny $1$}}
\def\2{\mbox{\tiny $2$}}
\def\3{\mbox{\tiny $3$}}
\def\4{\mbox{\tiny $4$}}
\def\5{\mbox{\tiny $5$}}
\def\6{\mbox{\tiny $6$}}
\def\7{\mbox{\tiny $7$}}
\def\8{\mbox{\tiny $8$}}
\def\9{\mbox{\tiny $9$}}
\def\min1{\mbox{\tiny $(-\,1)$}}
\def\m2{\mbox{\tiny $(-\,2)$}}
\def\={\mbox{\tiny $=$}}
\def\D{\mbox{\tiny D}}

\def\F{\mbox{\tiny F}}

\begin{document}
%

\title{\Large POTENTIAL SCATTERING IN DIRAC FIELD THEORY}

\author{
Stefano De Leo \inst{1}
\and Pietro Rotelli\inst{2} }

\institute{
Department of Applied Mathematics, University of Campinas\\
PO Box 6065, SP 13083-970, Campinas, Brazil\\
{\em deleo@ime.unicamp.br}\\
 \and
Department of Physics, University of
Salento and INFN, Lecce\\
PO BOX 193, CAP 73100, Lecce, Italy\\
{\em rotelli@le.infn.it} }


\date{Submitted: {\em March, 2009}. Revised: {\em April, 2009}. Accepted: {\em May, 2009}.}

\abstract{We develop the potential scattering of a spinor within
the context of perturbation field theory. As an application, we
reproduce, up to second order in the potential, the diffusion
results for a potential barrier of quantum mechanics. An immediate
consequence is a simple generalization to arbitrary potential
forms, a feature not possible in quantum mechanics.}


 \PACS{ {03.64.Nk}, {03.70.+k} {({\sc pacs}).}}













\titlerunning{\sc potential scattering in dirac field theory}

\maketitle


\section*{I. INTRODUCTION}

Quantum field theory  and the consequent renormalization theory
has had an unparallel success\cite{ZUB,GROSS,PES}. However, many
calculations in the physics literature are based upon one-particle
equations in quantum mechanics\cite{COHEN,SAK,FLU}. This is not
simply because quantum field theory calculations are, in general,
more cumbersome than quantum mechanical calculations. The real
reason is principally due to the method of calculation. Quantum
field theory calculations are often limited by the use of a
perturbation approach which can rarely be summed. For example,
quantum field theory successfully predicts the Lamb
shift\cite{LAMB} but, to the best of our knowledge, there is no
direct calculation of the hydrogen atom in field theory. Bound
states are almost exclusively treated in quantum
mechanics\cite{AMPL}. From quantum field theory one has, at best,
the two-body equations, based upon ladder diagrams, such as the
Bethe-Salpeter equation\cite{BeS1,BeS2}. It is even problematic to
prove the {\em mere} existence of some bound states in quantum
field theory\cite{SBSDR,DBSDR}, to discuss the space-time operator
in collision descriptions\cite{OR70,G81} or to analyze the
kinematics of spinning particle\cite{B1,B2,B3,RS98}. Another
example is that of potential scattering. Many important phenomena
such as tunneling\cite{HE,REC,TSDR,KRE1,KRE2,TDDR}, multiple
diffusion\cite{DSDR,DDDR}, Klein paradox\cite{K1,K2,K3,KREK,KDR},
to mention but a few, are described in quantum mechanics be they
with the Schr\"odinger, Klein-Gordon or Dirac equation.
 Must one be
limited for the description of these phenomena to quantum
mechanics? Are the quantum mechanical results exact? Generally,
experiment answers our doubts, but this is not always feasible.

In this paper, we intend to propose a field theoretic study of
potential scattering, one of the mainstays of quantum mechanics.
First, we write down the potential interaction term for the Dirac
lagrangian. This is followed by an outline of the quantum field
theory treatment of this system. We then apply the perturbation
series to a specific potential, that of a single one-dimensional
barrier, impinged upon perpendicularly by an incoming momentum
eigenstate. We calculate the lowest order contributions and
compare them with a Taylor expansion of the expressions for
transmission and reflection amplitudes for above barrier
diffusion, obtained with the Dirac equation\cite{DDDR}. We can in
this way determine if there are any discrepancies between quantum
mechanics and quantum field theory calculations and directly test
our quantum field theory approach.

In the next section, we present the quantum field theory
formalism\cite{PES}. In section III, we calculate the zeroth and
first order terms. In section IV we derive, in some detail, the
calculation of the second order terms. Our results are compared
with those of the Dirac equation  in section V. We conclude in
that section with a discussion of our results.

\section*{II. FORMALISM}

The first quantized quantum mechanical equation for a Dirac spinor
involving a time-independent one-dimensional potential $V(x_{\3})$
is given by
\begin{equation}
\label{eqd}
\left(i\partial\hspace*{-.19cm}\slash-m\right)\psi(x)=V(x_{\3})\,
\gamma_{\0}\,\psi(x)\,\,,
\end{equation}
where $x=(t,\boldsymbol{x})$.  Actually, the above choice $V\,
\gamma_{\0}\,\psi\,$ is for an {\em electrostatic} potential,
since it  can be derived from a four-vector potential interaction
 $A_{\mu} \gamma^{\mu}\,\psi$, when $\boldsymbol{A}=0$ and $A_{\0}=V$.
 An alternative, and
 generally non equivalent choice would be the {\em scalar} interaction
 $V\,\psi$. In standard field theory the two choices imply spinors
 interacting through the exchange of vector and scalar particles.
 However, in contrast to standard field theory, we shall not
 second-quantize this potential field. Throughout this paper $V$ will be a
 (c-number) function. This allows us to compare our results directly with
 those of the Dirac equation.

 From Eq.(\ref{eqd}), which is equally valid for the Dirac field, we
 immediately derive the following Lagrangian density
\begin{equation}
\label{lag}
\mathcal{L}(x)=\overline{\psi}(x)\left(i\partial\hspace*{-.19cm}\slash-m\right)
\psi(x)-V(x_{\3})\, \overline{\psi}(x) \gamma_{\0} \psi(x)\,\,.
\end{equation}
The spinor field can be quantized in the standard way. In the
interaction picture, the resulting spinor field takes the well
known form
\begin{equation}
\psi(x)=\int
\frac{\mbox{d}^{\3}\boldsymbol{k}}{(2\pi)^{^{\3}}2\,E(\boldsymbol{k})}\,
\sum_{s}\left[\,
a^{s}_{\boldsymbol{k}}\,u^s(\boldsymbol{k})\,e^{-ikx} +\,
b^s_{\boldsymbol{k}}\,v^{s}(\boldsymbol{k})\,e^{ikx}
\,\right]\,\,,
\end{equation}
where $k=[E(\boldsymbol{k}), \boldsymbol{k}]$,
$E(\boldsymbol{k})=\sqrt{\boldsymbol{k}^{^{2}}+m^{\2}}$ and
$k\,x=E(\boldsymbol{k})\,t -\boldsymbol{k}\cdot\boldsymbol{x}$ .
Technically, the time dependence of the field  is a consequence of
working in the interaction picture and the assumption of
anti-commutators for creation/annihilation operators.

The S-matrix for any process is formally given by
\begin{equation}
\label{sm} \langle \mbox{out}\,| \,S \,|\,\mbox{in}\rangle  =\,
_{\0}\hspace*{-0.04cm}\langle \mbox{out} \,|\,T\,\left\{\,\exp
\left[\,-\,i\,\int\,\mbox{d}^{\4}x\,\mathcal{H}_{int}(x)\right]\,\right\}
\,|\,\mbox{in}\rangle_{\0}\,\, ,
\end{equation}
where $T$ means time-ordered. In the above, it is understood that
only connected terms are  to be considered and the incoming and
outgoing states are defined by the action of creation operators
upon the free field vacuum state (indicated by the subscript $0$).
For an excellent derivation of the above we refer the reader to
the book of Peskin and Schroeder\cite{PES}. From Eq.(\ref{lag}),
we see that
\begin{equation}
\mathcal{H}_{int}(x)=V(x_{\3})\,\overline{\psi}(x)\gamma^{\0}\psi(x)\,\,
.
\end{equation}
 Eq.(\ref{sm}) can be evaluated by expanding in powers
 of the interaction Hamiltonian within the time-ordered
 integral. This is the basis of the perturbation approach.
 Normally, the unit ($0^{th}$ order) term is neglected and
 one considers only the so-called T-matrix. However, for our
 purposes we start by re-deriving this unit term since it must
 reproduce the transmission amplitude of the Dirac equation in the
 limit of {\em zero} potential.


 We choose the
 incoming state to be a  spinor of mass $m$
 travelling along the $x_{\3}$ axis,
       \[\boldsymbol{p}=(0,0,p_{\3})\,\,\,,\,\,\,
        E(\boldsymbol{p})=\sqrt{p_{\3}^{^{2}}+m^{\2}}:=E\,\,,\]
 with unspecified spin $s$.
 The outgoing state can only be a spinor of mass $m$,
 the only mass in the lagrangian, to which we assign an arbitrary spin
 $r$ and momentum $\boldsymbol{p'}$. As we shall prove in the following sections,
 only the values with $r=s$ and $\boldsymbol{p'}=(0,0,\pm \,p_{\3})$
 contribute, corresponding to spin conserving transmission and reflection.

\section*{III. ZEROTH AND FIRST ORDER CONTRIBUTIONS}

Our objective is to calculate in field theory the transmission and
reflection {\em amplitude} defined in quantum mechanics. This
involves some significant difference when compared to other field
theory calculations.

The first term in the expansion of Eq.(\ref{sm})  is simply given
by
\begin{equation}
_{\0}\langle \,
\boldsymbol{p'},r\,|\,\boldsymbol{p}\,,s\rangle_{\0} =
  2E\,(2\pi)^{^{3}}\,\delta^{^{3}}(\boldsymbol{p'}-\boldsymbol{p})\,\,
  \delta_{rs}\,\,.
\end{equation}
 This must correspond to the unit contribution of a
non-interacting quantum mechanical particle. Thus, we see that the
above result has to be integrated over the final three-momentum
$\boldsymbol{p'}$ and specifically with the standard
relativistically invariant measure,
\begin{equation}
\int \frac{\mbox{d}^{\3}\boldsymbol{p'}}{{(2\pi)^{^{\3}}}2\,E(
\boldsymbol{p'})}\,\,
_{\0}\langle\boldsymbol{p'},r\,|\,\boldsymbol{p}\,,s\rangle_{\0}=
\delta_{rs}\,\,.
\end{equation}
However, it is important to note that this results yields a
non-zero contribution only for $\boldsymbol{p'}=\boldsymbol{p}$,
i.e. for
 transmission, as required. The explicit Kronecker delta
 reminds us that there is no spin flip contribution. This delta is
 sometimes not written in the literature but expressed in words.
 In this paper we shall, instead, display it explicitly, even though
  it will be a common factor throughout. Summing over the final
  spin is seen not to be advisable since it would lose information.

 Consequent to the above discussion all our field theoretic terms
 will henceforth be integrated over the final state momentum,
 but not summed over final state spins.
 It is a little unusual to do final state momentum integrations
 before squaring the amplitude
 as occurs in the calculation of cross-sections and decay rates.
 However, a positive aspect is that it avoids the encounter
 of the square of delta functions and the subsequent need to define
  rates per unit time and unit volume. Actually, the final state
  momentum integration will never be completed. We will always
  separate (and not add) the transmitted and reflected
  contributions.
  Other unusual features compared to standard  field theory calculations is that
  the final state can only be a single outgoing particle and that
  three-momentum is not conserved because of reflection.

 For the  first order term, we must calculate
\begin{equation}
-\,i\,\,_{\0}\langle \, \boldsymbol{p'},r\,|\,T\left[\,\int
\mbox{d}^{\4}x\, V(x_{\3})\,
\bar{\psi}(x)\gamma^{\0}\psi(x)\right]\,|\,
\boldsymbol{p}\,,s\rangle_{\0}\,\,.
\end{equation}
The fields $\overline{\psi}$ and $\psi$ must be contracted with
the outgoing  and incoming states  yielding, after adding final
state momentum integration
\begin{equation}
 -\,i\int \frac{\mbox{d}^{\3}\boldsymbol{p'}}{(2\pi)^{^{\3}}2\,E(\boldsymbol{p'})}\,\,
 \mbox{d}^{\4}x\,
\,V(x_{\3})\, \langle0|u^{r\,\dag}(\boldsymbol{p'})\,e^{i(p'-p)x}
\,u^s(\boldsymbol{p})|0\rangle\,\,.
\end{equation}
 Performing the
space-time  integrals, with the exception for the coordinate
$x_{\3}$, and the transverse outgoing momentum integrals, we
obtain
\begin{equation}
 -\,i\int \frac{\mbox{d}p'_{\3}}{4\pi\,E}\,
 \mbox{d}x_{\3}\, V(x_{\3})\, \delta(E'-E)
\,\exp[-\,i\,(p'_{\3}-p_{\3})\,x_{\3}]\,
\langle0|u^{r\,\dag}(p'_{\3})\,u^s(p_{\3})|0\rangle \,\,,
\end{equation}
where $E'=\sqrt{{p_{\3}'}^{\2}+m^{\2}}$. For the integration over
the third component of the momentum, we use the identity
\[
\delta(E'-E) = \frac{E}{p_{\3}}\,\left[
\,\delta(p'_{\3}-p_{\3})+\delta(p'_{\3}+p_{\3})\,\right]\,\,,
\]
 Thus, in the amplitude there will be two incoherent
contributions. One for $p'_{\3}=p_{\3}$ (transmission) and one for
$p'_{\3}=-p_{\3}$ (reflection). These must be treated separately.
For transmission, we find
\begin{equation}
-\,\frac{i}{2\,p_{\3}}\,\,
\langle0|u^{r\,\dag}(p_{\3})\,u^s(p_{\3})|0\rangle\, \int
\mbox{d}x_{\3}\, V(x_{\3}) = -\,i\,\frac{E}{p_{\3}}\,\, \int
\mbox{d}x_{\3}\, V(x_{\3})\,\,\delta_{rs} \,\,.
\end{equation}
While, for reflection, we obtain
\begin{equation}
 -\,\frac{i}{2\,p_{\3}}\,\,
\langle0|u^{r\,\dag}(-p_{\3})\,u^s(p_{\3})|0\rangle\, \int
\mbox{d}x_{\3}\, V(x_{\3})\, \exp[2\,i\,p_{\3}x_{\3}]=
-\,i\,\frac{m}{p_{\3}}\,\,  \int \mbox{d}x_{\3}\, V(x_{\3})\,
\exp[2\,i\,p_{\3}x_{\3}] \,\delta_{rs}\,\,.
\end{equation}
We must now perform
the $x_{\3}$ integrals. To do so, we must now select a specific
potential. We apply our formalism to a \emph{barrier} potential of
height $V_{\0}$ and situated on the $x_{\3}$ axis between 0 and L,
\[
V(x_{\3})=\{\,0\,\,\,\,\,(\,x_{\3}<0\,)\,\,\,,\,\,\,\,\,V_{\0}\,\,\,\,\,
(\,0<x_{\3}<L\,)\,\,\,,\,\,\,\,\,
0\,\,\,\,\,(\,x_{\3}>L\,)\,\}\,\,.
\]
The results of the elementary $x_{\3}$ integrations are
\begin{equation}
-\,i\,\frac{E}{p_{\3}}\,V_{\0}\,L\,\delta_{rs}
\end{equation}
for transmission, and
\begin{equation}
-\,i\,\frac{m\,V_{\0}}{p_{\3}^{^{2}}}\,
\sin(p_{\3}L)\,\exp[\,i\,p_{\3}L]\,\delta_{rs}
\end{equation}
for reflection.

A comment before proceeding to the calculation of the second order
terms. Because of the $x_{\3}$ dependence of the potential $V$, we
cannot perform all of the space-time integrals trivially, and thus
we do \emph{not} have overall energy-momentum conservation.
However, energy conservation (a consequence of the $x_{\0}$
integration) together with the mass shell conditions implies the
conservation of the square of the three momentum. This means that
the allowed contributions are  transmission and reflection. There
is nothing unusual in this. Even in reflection from a step
function only momentum squared is conserved as also for our bound
state quantum mechanical problem.

 \section*{IV. SECOND ORDER CONTRIBUTION}

The second order contribution is given by
\begin{equation}
-\,\mbox{$\frac{1}{2}$}\,\,_{\0}\langle \,
\boldsymbol{p'},r\,|\,T\left[\,\int \mbox{d}^{\4}x\,
\mbox{d}^{\4}y\, V(x_{\3})\,V(y_{\3})\,
\overline{\psi}(x)\gamma^{\0}\psi(x)
\overline{\psi}(y)\gamma^{\0}\psi(y)\,\right]\,|\,
\boldsymbol{p}\,,s\rangle_{\0}\,\,.
\end{equation}
The two modes of contraction result in identical contributions and
hence simply cancel the factor of $\frac{1}{2}$ above. We are thus
left with the calculation of
\begin{equation}
-\,\int
\frac{\mbox{d}^{\3}\boldsymbol{p'}}{(2\pi)^{^{\3}}2E(\boldsymbol{p'})}\,
\mbox{d}^{\4}x\,\mbox{d}^{\4}y\, V(x_{\3})\,V(y_{\3})\,
\langle0|u^{r\,\dag}(\boldsymbol{p'})\,e^{ip'x}\,S_{\F}(x-y)\,
e^{-ipy}\,\gamma_{\0}\, \,u^{s}(\boldsymbol{p})|0\rangle\,\,,
\end{equation}
where  $S_{\F}(x-y)$ is the Feynman propagator,
\[
S_{\F}(x-y) =\int
\frac{\mbox{d}^{\4}k}{(2\pi)^{^{4}}}\,\frac{i\,(k\hspace*{-.18cm}/
+m)}{k^{^2}-m^{\2}+i\epsilon}\,e^{-ik(x-y)}\,\,.
\]
Before continuing, we wish to observe that there are eight
space-time integrals to perform and seven momentum integrals. Only
six space time integrals can be immediately executed yielding six
delta functions. These can then be used to integrate over momenta
and specifically over all but the $k_{\3}$ momentum. Also
remaining are the $\mbox{d}x_{\3}$ and $\mbox{d}y_{\3}$ integrals.
The $k_{\3}$ momentum will be integrated in the complex plane and
the final space integrals by elementary means. During this
procedure, we use the treatment described in the previous section
for the $\delta(E'-E)$ to perform the $\mbox{d}p_{\3}'$ integral.
This is the point in which separation between transmission and
reflection occurs.

The integration over the space-time variables, excluding $x_{\3}$
and $y_{\3}$, yields the following delta functions
\[
\delta(p_{\0}'-k_{\0})\,\delta(p_{\1}'-k_{\1})\,\delta(p_{\2}'-k_{\2})\,
\delta(p_{\0}-k_{\0})\,\delta(p_{\1}-k_{\1})\,\delta(p_{\2}-k_{\2})\,\,.
\]
We then integrate over the transverse momenta both for
$\boldsymbol{k}$ and $\boldsymbol{p'}$. The integral over
$\mbox{d}k_{\0}$ leaves us with one remaining energy delta
function which we rewrite as a sum of delta's in $p'_{\3}$. One
delta for transmission and one for reflection. After performing
the $\mbox{d}p_{\3}'$ integral, for transmission, we obtain
\begin{eqnarray}
-\,\int \frac{\mbox{d}k_{\3}}{4\pi \,p_{\3}}\,
\mbox{d}x_{\3}\,\mbox{d}y_{\3}\, V(x_{\3})\,V(y_{\3})\,\exp[i
p_{\3}(y_{\3} - x_{\3})]\,\, \times \nonumber \\
 \langle0|u^{r\,\dag}(p_{\3}) \,\,
  \frac{i\,(\gamma_{\0}E -\gamma_{\3}k_{\3}+m)}{p_{\3}^{^2}-k_{\3}^{^{2}}+
  i \epsilon} \,
\exp[ik_{\3} (x_{\3}- y_{\3})]\,\gamma_{\0}\,u^s(p_{\3})|0\rangle
\end{eqnarray}
and for reflection,
\begin{eqnarray}
-\,\int \frac{\mbox{d}k_{\3}}{4\pi \,p_{\3}}\,
\mbox{d}x_{\3}\,\mbox{d}y_{\3}\, V(x_{\3})\,V(y_{\3})\,\exp[i
p_{\3}(y_{\3} + x_{\3})]\,\, \times \nonumber \\
 \langle0|u^{r\,\dag}(-p_{\3}) \,\,
  \frac{i\,(\gamma_{\0}E -\gamma_{\3}k_{\3}+m)}{p_{\3}^{^2}-k_{\3}^{^{2}}+
  i \epsilon} \,
\exp[ik_{\3} (x_{\3}- y_{\3})]\,\gamma_{\0}\,u^s(p_{\3})|0\rangle
\end{eqnarray}
The $\mbox{d}k_{\3}$ integral can be performed with the help of
the residue theorem.  However, the result is different for
$x_{3}>y_{3}$ or $x_{3}<y_{3}$. According to the case, we must
close the contour above or below the axis.
\begin{eqnarray*}
\int \mbox{d}k_{\3} \,
  \frac{i\,(\gamma_{\0}E -\gamma_{\3}k_{\3}+m)}{p_{\3}^{^2}-k_{\3}^{^{2}}+
  i \epsilon} \,
\exp[ik_{\3} (x_{\3}- y_{\3})] & = & \theta(x_{\3}-y_{\3})\,
\frac{\pi}{p_{\3}}\,(\gamma_{\0}E
-\gamma_{\3}p_{\3}+m)\,\exp[ip_{\3} (x_{\3}- y_{\3})] + \\
 &  & \theta(y_{\3}-x_{\3})\, \frac{\pi}{p_{\3}}\,(\gamma_{\0}E
+\gamma_{\3}p_{\3}+m)\,\exp[ip_{\3} (y_{\3}-x_{\3})]\,\,.
\end{eqnarray*}
 Whence, the second order contribution to transmission, $p'_{\3}=p_{\3}$, reads
\begin{eqnarray}
-\,\frac{1}{4\,p_{\3}^{^{2}}}\, \int \,
\mbox{d}x_{\3}\,\mbox{d}y_{\3}\, V(x_{\3})\,V(y_{\3})\,\exp[i
p_{\3}(y_{\3} - x_{\3})]\,\times  & & \nonumber \\
\left\{\,\theta(x_{\3}-y_{\3})\,\langle0|u^{r\,\dag}(p_{\3};E)
\,(\gamma_{\0}E -\gamma_{\3}p_{\3}+m)
\,\gamma_{\0}\,u^s(p_{\3};E)|0\rangle \right. \,\exp[ip_{\3}
(x_{\3}-
y_{\3})] \,\,+ & &   \nonumber  \\
\left.\,\theta(y_{\3}-x_{\3})\,\langle0|u^{r\,\dag}(p_{\3};E)
\,(\gamma_{\0}E
+\gamma_{\3}p_{\3}+m)\,\gamma_{\0}\,u^s(p_{\3};E)|0\rangle \,
\exp[ip_{\3} (y_{\3}- x_{\3})] \,
\right\} & = & \nonumber \\
-\,\frac{1}{p_{\3}^{^{2}}}\, \int \,
\mbox{d}x_{\3}\,\mbox{d}y_{\3}\, V(x_{\3})\,V(y_{\3})\,
\left\{\,\theta(x_{\3}-y_{\3})\,E^{^{2}} +
\theta(y_{\3}-x_{\3})\,m^{\2}\,\exp[2\,ip_{\3} (y_{\3}-
x_{\3})]\,\right\}\,\delta_{rs} & . &
\end{eqnarray}
While, the second order contribution  for reflection,
$p'_{\3}=-p_{\3}$, reads
\begin{eqnarray}
-\,\frac{1}{4\,p_{\3}^{^{2}}}\, \int \,
\mbox{d}x_{\3}\,\mbox{d}y_{\3}\, V(x_{\3})\,V(y_{\3})\,\exp[i
p_{\3}(y_{\3} + x_{\3})]\,\times & & \nonumber \\
\left\{\,\theta(x_{\3}-y_{\3})\,\langle0|u^{r\,\dag}(-p_{\3};E)
\,(\gamma_{\0}E
-\gamma_{\3}p_{\3}+m)\,\gamma_{\0}\,u^s(p_{\3};E)|0\rangle
\right.\, \exp[ip_{\3} (x_{\3}- y_{\3})] \, \, + & & \nonumber  \\
\left.\,\theta(y_{\3}-x_{\3})\,\langle0|u^{r\,\dag}(-p_{\3};E)
\,(\gamma_{\0}E
+\gamma_{\3}p_{\3}+m)\,\gamma_{\0}\,u^s(p_{\3};E)|0\rangle\,
\exp[ip_{\3} (y_{\3}- x_{\3})] \, \right\} & = &
\nonumber\\
-\,\frac{m\,E}{p_{\3}^{^{2}}}\, \int \,
\mbox{d}x_{\3}\,\mbox{d}y_{\3}\, V(x_{\3})\,V(y_{\3})\,
\left\{\,\theta(x_{\3}-y_{\3})\, \exp[2\,ip_{\3} x_{\3}]\, +
\theta(y_{\3}-x_{\3})\,\exp[2\,ip_{\3}y_{\3}]\,\right\}\,\delta_{rs}
& . &
\end{eqnarray}
After performing the $x_{\3}$ integration, we find
\begin{equation}
-\,\left\{\frac{E^{^{2}}L^{^{2}}}{2\,p_{\3}^{^{2}}} +
 \frac{m^{\2}}{4\,p_{\3}^{^{4}}}\, \left( 1 - \exp[\,2\,i\,p_{\3}L]+ 2\,i\,p_{\3}L
 \right)
 \right\}\,
 V_{\0}^{^{2}}\,\delta_{rs}\,\,,
 \end{equation}
for transmission, and
\begin{equation}
 i\,\frac{m\,E}{p_{\3}^{^{4}}}\,\exp[\,ip_{\3}L]\,
 \left\{p_{\3}L \,\exp[\,ip_{\3}L] -\sin(p_{\3}L)\right\}\, V_{\0}^{^{2}}\,
 \delta_{rs}\,\,,
 \end{equation}
for reflection.

\section*{V. CONCLUSIONS}

  In the previous two sections, we have calculated the $0^{th},
  1^{st}$  and  $2^{nd}$ order contributions of spinor field theory to
  scattering  off a simple one-dimensional barrier. Now, we
  compare these results with those of quantum mechanics.
  We recall the full expressions for the Dirac transmission and reflection
  coefficients\cite{DDDR},
\begin{eqnarray}
R_{\D} & = & -\,i\,
\frac{m\,V_{\0}}{q_{\3}p_{\3}}\,\,\sin(q_{\3}L)\,\mbox{\Large
$/$}\left[\,\cos(q_{\3}L)\,-\,i\,\frac{p_{\3}^{\2}-E\,
V_{\0}}{q_{\3}\,p_{\3}}\,
\sin(q_{\3}L)\,\right]\,\,, \nonumber \\
T_{\D} & = & \exp[\,-\,ip_{\3} L] \,\mbox{\Large
$/$}\left[\,\cos(q_{\3}L)\,-\,i\,\frac{p_{\3}^{\2}-E\,V_{\0}}{q_{\3}\,p_{\3}}\,
\sin(q_{\3}L)\,\right]\,\,,
\end{eqnarray}
where $q_{\3}=\sqrt{(E-V_{\0})^{^{2}}-m^{\2}}$. We have
re-expressed $R_{\D}$ in order to explicitly display the $V_{\0}$
dependence. Polarization conservation in \cite{DDDR} was stated
and not given in terms of a Kroneker delta. The first terms of the
Taylor expansions in $V_{\0}$ are
\begin{eqnarray}
R_{\D} & \approx &
 -\,i\,\frac{m\,V_{\0}}{p_{\3}^{^{2}}}\,\sin(p_{\3}L)\,\exp[\,ip_{\3} L]+
 i\,\frac{m\,E\,V_{\0}^{^{2}}}{p_{\3}^{^{4}}}\,\exp[\,ip_{\3} L]\,
 \left\{p_{\3}L \,\exp[\,ip_{\3} L] -\sin(p_{\3}L)\right\}
 \,\,, \nonumber \\
T_{\D} & = & 1 -\,i\,\frac{E}p_{\3}\,V_{\0}\,L -
 \left[\frac{E^{^{2}}L^{^{2}}}{2\,p_{\3}^{^{2}}} +
 \frac{m^{\2}}{4\,p_{\3}^{^{4}}}\, \left( 1 - \exp[\,2\,ip_{\3} L]+ 2\,ip_{\3}L \right)
 \right]\,
 V_{\0}^{^{2}}\,\,.
\end{eqnarray}
Agreement with our field theory calculations up to and including
second order is found. This result was not an obvious prediction,
a priori. First, because the calculational methods are very
different. Secondly, because while in quantum mechanics continuity
plays a major role, in field theory continuity never appears, but
antiparticles do, for example they are intrinsic in the Feynman
propagator. Furthermore, while for diffusion there are no
particular surprises with the quantum mechanical results, this is
not true for tunnelling. In tunnelling, it has been shown
elsewhere that the Hartman effect exists. This is an apparent
violation of causality which we do not expect in any field theory
calculation. Indeed, it was one of the original stimuli for
developing the above formalism. Unfortunately, our results do not
extend into the tunnelling region. However, one of the advantages
of the above agreement, which we postulate holds to {\em all}
orders, is that it confirms the correctness of our procedures.

An important consequence of our formalism is that since the
potential functions are merely integrated over, the application of
our procedures to {\em any} potential shape is straightforward,
and in some cases, when the integrations can be performed
analytically, even simple. This is not the case for general
step-wise potentials in quantum mechanics. There, the
calculational difficulties (coupled matrix equations) grows at
least linearly with the number of potential discontinuities.

 The
perturbation approach is  not valid for tunneling. This merits a
comment. In quantum mechanics, it is possible to treat tunneling
and diffusion simultaneously. The only difference, which can be
left to almost the end of the calculation, is that the momentum
$q_{\3}$ is either real (for diffusion) or pure imaginary (for
tunneling). This essentially transforms trigonometric functions in
hyperbolic functions. However, the problem lies elsewhere. It is
that while a factor like $\sqrt{(E-V_{\0})^{^{2}}-m^{\2}}$ for
$E>V_{\0}+m$ (diffusion) can be Taylor expanded in $V_{\0}$, for
$E<V_{\0}-m$ (tunneling) it {\em cannot}. Of course, if one
believes that the tunneling results are an analytic continuation
of the diffusion results, our field theory calculation should
yield the same tunneling results. However, a place where this is
clearly not the case is in the Klein energy zone. The Klein
paradox for a step potential is interpreted as the creation at the
potential discontinuity of particle-antiparticle pairs. This is
obviously inconsistent with the single particle nature of the
Dirac equation, but has been seen as an anticipation of field
theory. The characteristic of this paradox is a reflection
coefficient greater than one. If one treats the barrier potential,
in the Klein zone, in the standard way, one finds instead {\em no}
paradox. Recently, a pair creation treatment of the barrier has
been provided\cite{KDR}. It implies the existence of a new type of
spatial localization, since the antiparticles created at each
reflection are necessarily blocked within the barrier, which they
see as a potential well. So, what more appropriate a subject for a
field theoretical treatment than the Klein zone of the barrier. At
the moment, this remains a feature objective.

\end{document}